\documentclass[12pt,preprint]{aastex}

\newcommand{\ms}{\mbox{m s$^{-1}~$}}

\newcommand{\kms}{\mbox{km s$^{-1}~$}}
\newcommand{\msun}{M$_{\odot}$\ }

\newcommand{\mjup}{M$_{\rm JUP}$}

\newcommand{\msini}{$M \sin i~$}
\newcommand{\vsini}{$V \sin i~$}

\shortauthors{Carter et al.}
\shorttitle{HD 70642}
\received{}
\accepted{}
\begin{document}

\title{A Planet in a Circular Orbit with a 6 Year Period$~^{1}$}

\author{Brad D. Carter\altaffilmark{2},
R. Paul Butler\altaffilmark{3}, 
C. G. Tinney\altaffilmark{4},
Hugh R. A. Jones\altaffilmark{5},
Geoffrey W. Marcy\altaffilmark{6},
Chris McCarthy\altaffilmark{3},
Debra A. Fischer\altaffilmark{6},
Alan J. Penny\altaffilmark{7}}

\email{carterb@usq.edu.au}

\altaffiltext{1}{Based on observations obtained at 
the Anglo--Australian Telescope, Siding Spring, Australia}

\altaffiltext{2}{Faculty of Sciences, University of Southern Queensland,
Toowoomba, Queensland 4350, Australia}

\altaffiltext{3}{Department of Terrestrial Magnetism, Carnegie Institution
of Washington, 5241 Broad Branch Road NW, Washington D.C. USA 20015-1305}

\altaffiltext{4}{Anglo--Australian Observatory, P.O. Box 296,
Epping, NSW 1710, Australia}

\altaffiltext{5}{Astrophysics Research Institute, Liverpool John Moores
University, Twelve Quays House, Egerton Wharf, Birkenhead CH41 1LD, UK}

\altaffiltext{6}{Department of Astronomy, University of California,
Berkeley, CA USA  94720 and at Department of Physics and Astronomy,
San Francisco State University, San Francisco, CA, USA 94132}

\altaffiltext{7}{Rutherford Appleton Laboratory,
Chilton, Didcot, Oxon, OX11 0QX, UK}

\begin{abstract}

Precision Doppler velocity measurements from the 3.9--m
Anglo-Australian Telescope reveal a planet with a 6 year
period orbiting the G5 dwarf HD 70642. The a = 3.3 AU orbit
has a low eccentricity (e = 0.1), and the minimum 
(\msini) mass of the planet is 2.0 \mjup.  The host star is
metal rich relative to the Sun, similar to most stars with
known planets.  The distant and approximately circular orbit of this 
planet makes it a member of a rare group to emerge from precision 
Doppler surveys.

\end{abstract}


\section{Introduction}
\label{intro}

Of the 77 extrasolar planets currently listed by the IAU
Working Group on Extrasolar
Planets\footnote{$http://www.ciw.edu/boss/IAU/div3/wgesp/planets.shtml$}
(including planet candidates published in a refereed journals with
\msini $<$ 10 \mjup), only three systems have been found to harbor 
planets in circular orbits
(e $<$ 0.1) orbits beyond 0.5 AU -- 47 UMa (Fischer et al. 2002;
Butler \& Marcy 1996), HD 27442 (Butler et al. 2001), and HD 4208
(Vogt et al. 2002).  With 13 ``51 Peg--type'' planets (P $<$ 5 d),
and $\sim$60 eccentric planets (e $>$ 0.1), the long period circular
orbits are the rarest of the three types of planetary systems to
emerge over the last 8 years.

With one exception, all the IAU Working Group List planets orbit
within 4 AU of their host stars.  As all these planets have been
discovered via the precision Doppler technique, there is a strong
selection bias toward discovering systems with small semimajor axes.
Unsurprisingly, the only extrasolar planet so far found to orbit
beyond 4 AU was detected by the precision Doppler survey that has
been gathering data the longest (Marcy et al. 2002).

Perhaps the most critical question facing the field of extrasolar
planetary science is ``Are Solar System analogs (ie. systems with
giants planets in circular orbits beyond 4 AU and small rocky planets
orbiting in the inner few AU) ubiquitous, or rare?'' Existing precision
Doppler surveys will become sensitive to giant planets orbiting
beyond 4 AU by the end of this decade, though only those programs with
long term precision of 3 \ms or better will be able to determine if the
orbits of such planets are eccentric or circular (Butler et al.
2001, Figure 11).

We report here a new extrasolar planet in an approximately circular orbit
beyond 3 AU, discovered with the 3.9m Anglo--Australian Telescope (AAT).
The Anglo-Australian Planet Search program is described in Section 2.
The characteristics of the host star and the precision Doppler
measurements are presented in Section 3.  A discussion follows.

\section{The Anglo--Australian Planet Search}
\label{obs}

The Anglo-Australian Planet Search began in 1998 January, and is
currently surveying 250 stars.  Fourteen planet candidates with
\msini ranging from 0.2 to 10 \mjup \ have first been published 
with AAT data (Tinney et al. 2001; Butler et al. 2001; Tinney et
al. 2002a; Jones et al. 2002a; Butler et al. 2002; Jones et al.
2002b; Tinney et al. 2003a; Jones et al. 2003), and an additional
four planet candidates have been confirmed with AAT data
(Butler et al. 2001). 

Precision Doppler measurements are made with the University College 
London Echelle Spectrograph (UCLES) (Diego et al. 1990).  An iodine 
absorption cell
(Marcy \& Butler 1992) provides wavelength calibration from
5000 to 6000 \AA.  The spectrograph PSF and wavelength calibration
are derived from the embedded iodine lines (Valenti et al. 1995;
Butler et al. 1996).  This system has demonstrated long term
precision of 3 \ms (Butler et al. 2001), similar to (if not 
exceeding) the iodine systems on the Lick 3-m (Butler et al. 1996;
1997) and the Keck 10-m (Vogt et al. 2000).

\section{HD 70642}
\label{results}
 
HD 70642 (HIP 40952, SAO 199126) is a nearby G5 dwarf, at a distance of
28.8 pc (Perryman et al. 1997), a $V$ magnitude of 7.17, and an absolute
magnitude of $M_{\rm V}$ = 4.87.  The star is photometrically stable within
Hipparcos measurement error (0.01 magnitudes).  The star is chromospherically
inactive, with log$R$'(HK) $=$ $-$4.90 $\pm$0.06, determined from AAT/UCLES
spectra of the Ca II H\&K lines (Tinney et al. 2003b; Tinney et al. 2002b).
Figure 1 shows the H line compared to the Sun.  The chromospherically
inferred age of HD 70642 is $\sim$4 Gyr.

Spectral synthesis (LTE) of our AAT/UCLES spectrum of HD 70642 yields
T$_{eff}$ $=$5670 $\pm$20 K and \vsini $=$ 2.4 $\pm$1 \kms consistent
with its status as a middle--aged G5 dwarf.  Like most planet
bearing stars, HD 70642 is metal rich relative to the Sun.  We
estimate [Fe/H] $=$ $+$0.16 $\pm$0.02 from spectral synthesis, in
excellent agreement with the photometric determination of Eggen
(1998).  While Ni tracks Fe for most G \& K dwarfs, the [Ni/H] $=$
$+$0.22 $\pm$0.03 appears slightly high for HD 70642.  The mass of
HD 70642 estimated from $B$--$V$, M$_{\rm Bol}$, and [Fe/H] is 1.0
$\pm$0.05 \msun.

A total of 21 precision Doppler measurements of HD 70642 spanning more 
than 5 years are listed in Table 1 and shown in Figure 2.  The solid line 
in Figure 2 is the best--fit Keplerian.  The Keplerian parameters are
listed in Table 2.  The reduced $\chi_{\nu}^2$ of the Keplerian fit is
1.4.  Figure 3 is a plot of orbital eccentricity vs. semimajor axis for 
the planet orbiting HD70642, for extrasolar planets listed by the IAU Working 
Group on Extrasolar Planets, and Solar System planets out to 
Jupiter. HD 70642b joins 47 UMa c (Fischer et al. 2002) as the only planets
yet found in an approximately circular (e $\le$ 0.1) orbit beyond 3 
AU.

\section{Discussion}

Prior to the discovery of extrasolar planets, planetary systems were
predicted to be architecturally similar to the Solar System (Lissauer
1995; Boss 1995), with giant planets orbiting beyond 4 AU in circular
orbits, and terrestrial mass planets inhabiting the inner few AU.  The
landscape revealed by the first $\sim$80 extrasolar planets is quite
different.  Extrasolar planetary systems have proven to be much more
diverse than imagined, as predicted by Lissauer (1995), ``The variety
of planets and planetary systems in our galaxy must be immense and
even more difficult to imagine and predict than was the diversity of
the outer planet satellites prior to the Voyager mission.''

The discovery here of a Jupiter--mass planet in a circular orbit
highlights the existence, but also the rarity, of giant planets that
seem similar to the original theoretical predictions.  Review of all
the known extrasolar planets, both those described in refereed,
published journals
($http://www.ciw.edu/boss/IAU/div3/wgesp/planets.shtml$) and those in
the larger list of claimed extrasolar planets
($http://exoplanets.org$) shows that $\sim$7\% of extrasolar planets
orbiting beyond 0.5 AU reside in circular orbits ($e <$0.1).  Further
detections of planets beyond 1 AU are needed to determine if circular
orbits are more common for planets that orbit farther from the host
star.

Over the next decade precision Doppler programs will continue
to be the primary means of detecting planets orbiting stars
within 50 parsecs.  By the end of this decade, Doppler programs
carried out at precisions of 3 \ms or better by our group, and
by others (e.g., Mayor \& Santos 2002), will be sensitive to Jupiter
and Saturn--mass planets orbiting beyond 4 AU.  The central looming
question is whether these planets will commonly be found in circular
orbits, or if the architecture of the Solar System is rare.

Of the greatest anthropocentric interest are planets in intrinsically
circular orbits, as opposed to the short period planets in tidally
circularized orbits.  NASA and ESA have made plans for new telescopes
to detect terrestrial mass planets.  Transit missions such as COROT,
Kepler and Eddington may be sensitive to terrestrial mass planets
orbiting near 1 AU, providing valuable information on the incidence
of such planets.  As terrestrial planets make photometric signatures
of 1 part in 10,000, these missions may be subject to interpretive
difficulties that already challenge current ground--based transit searches
for Jupiter--sized planets.  Transit missions cannot determine orbital
eccentricity, and thus cannot determine if planets are Solar System
analogs.  These space--based transit missions are targeting stars at
several hundred parsecs, making follow--up by other techniques difficult.

Ground--based interferometric astrometry programs at Keck and VLT are
projected to begin taking data by the end of this decade.  These
programs are complementary to existing precision Doppler velocity
programs in that they are most sensitive to planets in distant orbits.
Like Doppler velocities, astrometry needs to observe one or more
complete orbits to make a secure detection and solve for orbital
parameters.  It is thus likely that the first significant crop of
ground--based astrometry planets will emerge after 2015.

The NASA Space Interferometry Mission (SIM) is scheduled to launch in
2009.  A key objective of the SIM mission is the detection of 
planets as small as 3 Earth--masses in 1 AU orbits around the nearest stars.
SIM offers the promise of determining actual masses of terrestrial
planets, thereby securing their status unambiguously. 
Simulations based on the SIM measurement specifications, along with the
proposed target stars, the 5 year mission lifetime, and a planet
mass function extrapolated to the Earth--mass regime\footnote{A 
power--law extrapolation admittedly fraught with uncertainty.}, 
yield predictions that as many as $\sim$5 terrestrial
planets could be found (Ford \& Tremaine 2003).  Giant planets
orbiting 2--5 AU from the host stars are also detectable with SIM,
though the orbital parameters will not be not well determined in a
5 year mission.  A 10 year SIM mission yields significantly better
orbital determination for Jupiter--analogs.  Overall the detection
capabilities of SIM are similar to existing precision Doppler programs
with a precision of 3 \ms (Ford \& Tremaine 2003), thereby providing
confirmation of known planets and unambiguous masses.

Direct imaging missions such as the NASA Terrestrial Planet Finder
(TPF) and the ESA DARWIN mission have the primary goal of detecting 
Earth--like
planets and obtaining low resolution spectra that might reveal
biomarkers.  Such missions will not return dynamical information and
hence will not directly reveal the masses of detected planets.
Current plans call for the launching of such missions around 2015,
perhaps optimistically.  We expect that continued Doppler
measurements, as well as future astrometric missions, will contribute
significantly to the interpretation of the unresolved companions
detected by TPF/DARWIN.

\acknowledgements

We acknowledge support by NSF grant AST-9988087, NASA grant
NAG5-12182, and travel support from the Carnegie Institution
of Washington (to RPB), NASA grant NAG5-8299 and NSF grant
AST95-20443 (to GWM), and by Sun Microsystems.  We also wish to
acknowledge the support of the NASA Astrobiology Institute.
We thank the Australian (ATAC) and UK (PATT) Telescope assignment
committees for allocations of AAT time.  We are grateful
for the extraordinary support we have received from the
AAT technical staff  -- E. Penny, R. Paterson, D. Stafford,
F. Freeman, S. Lee, J. Pogson, and G. Schaffer.  We would
especially like to express our gratitude to the AAO Director,
Brian Boyle.  The AAT Planet Search Program was created and
thrived under Brian's tenure as Director in large measure
because of his critical support and encouragement.  The Australia
Telescope National Facility (ATNF) is fortunate to have
Brian as their new Director.

\clearpage

\begin{deluxetable}{rrr}
\tablenum{1}
\tablecaption{Velocities for HD 70642}
\label{vel70642}
\tablewidth{0pt}
\tablehead{
\colhead{JD}           &    \colhead{RV}         & \colhead{Uncertainty} \\
\colhead{($-$2450000)}   &  \colhead{(m s$^{-1}$)} & \colhead{(m s$^{-1}$)} 
}
\startdata
   830.1082  &   -25.8  &  4.2 \\
  1157.2263  &   -36.4  &  4.4 \\
  1213.1051  &   -42.6  &  4.3 \\
  1236.0850  &   -37.5  &  5.2 \\
  1630.0095  &   -15.9  &  3.6 \\
  1717.8810  &    -9.4  &  3.9 \\
  1920.1348  &    15.0  &  4.8 \\
  1983.9687  &    13.7  &  5.8 \\
  2009.0210  &    13.2  &  4.6 \\
  2060.8744  &    27.7  &  3.4 \\
  2420.9072  &    12.7  &  3.1 \\
  2424.8981  &    11.4  &  3.1 \\
  2455.8416  &    20.0  &  2.8 \\
  2592.2229  &    12.7  &  2.9 \\
  2595.2255  &    15.2  &  3.4 \\
  2710.0700  &     0.8  &  3.0 \\
  2744.9571  &     0.3  &  3.1 \\
  2747.9155  &    -4.2  &  2.7 \\
  2749.9755  &    -7.4  &  2.2 \\
  2751.9384  &    -5.6  &  2.4 \\
  2785.9082  &    -3.4  &  2.4 \\
\enddata
\end{deluxetable}

\begin{deluxetable}{lcc}
\tablenum{2}
\tablecaption{Orbital solution for HD 70642}
\label{orbit}
\tablewidth{0pt}
\tablehead{
\colhead{Parameter} & \colhead{Value} & \colhead{Uncertainty}
}
\startdata
Orbital period $P$ (days) &   2231 & 400 \\
Velocity semiamplitude $K$ (m\,s$^{-1}$) & 32  & 5 \\
Eccentricity $e$ & 0.10 & 0.06\\
Periastron date (Julian Date) & 2451749  & 300 \\
$\omega$ (degrees) &  277 & 75 \\
M$\sin i$ (\mjup) & 2.0 \\
semimajor axis (AU) & 3.3 \\ 
N$_{\rm obs}$ & 21 \\
RMS (m\,s$^{-1}$) & 4.0 \\
\enddata
\end{deluxetable}

\clearpage

\begin{figure}
\epsscale{0.8}
\plotone{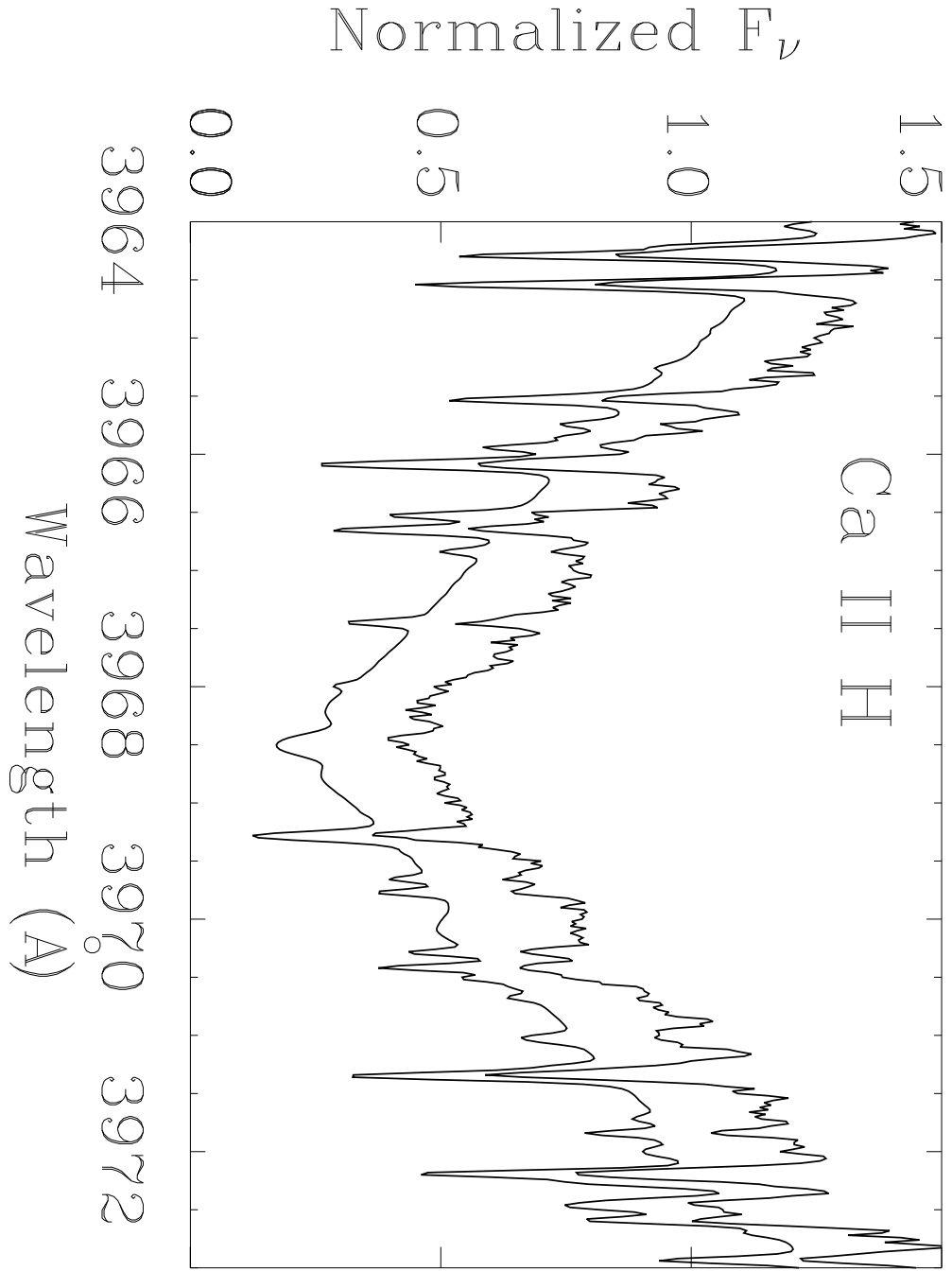}
\caption{Ca II H line of HD 70642 (top) compared to the Sun.  HD 70642
is a near solar twin in both mass and chromospheric activity.
Chromospherically quiescent stars such as these are intrinsically
stable at the 3 \ms level.}
\label{HD70642_H_line}
\end{figure}

\clearpage

\begin{figure}
\epsscale{0.8}
\plotone{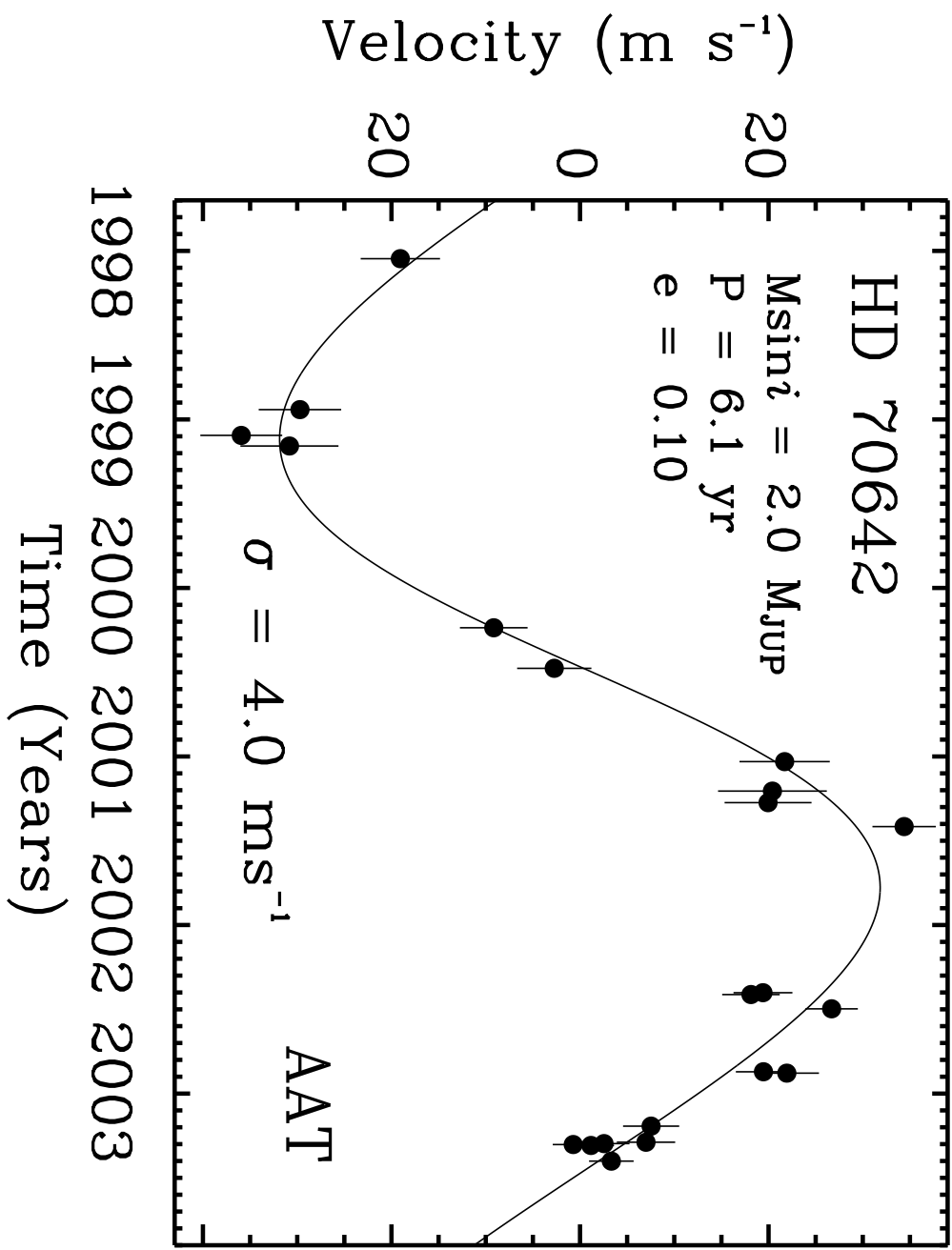}
\caption{Doppler velocities for HD 70642. 
The best--fit Keplerian is shown as a solid line, with period, P = 6.1 yr,
semiamplitude, K = 32 \ms, eccentricity e = 0.10, yielding \msini = 2.0 
\mjup. The RMS to the Keplerian fit, 4.0 \ms, is consistent with 
measurement
uncertainty.}
\label{HD70642_Velocities}
\end{figure}

\clearpage

\begin{figure}
\epsscale{0.8}
\plotone{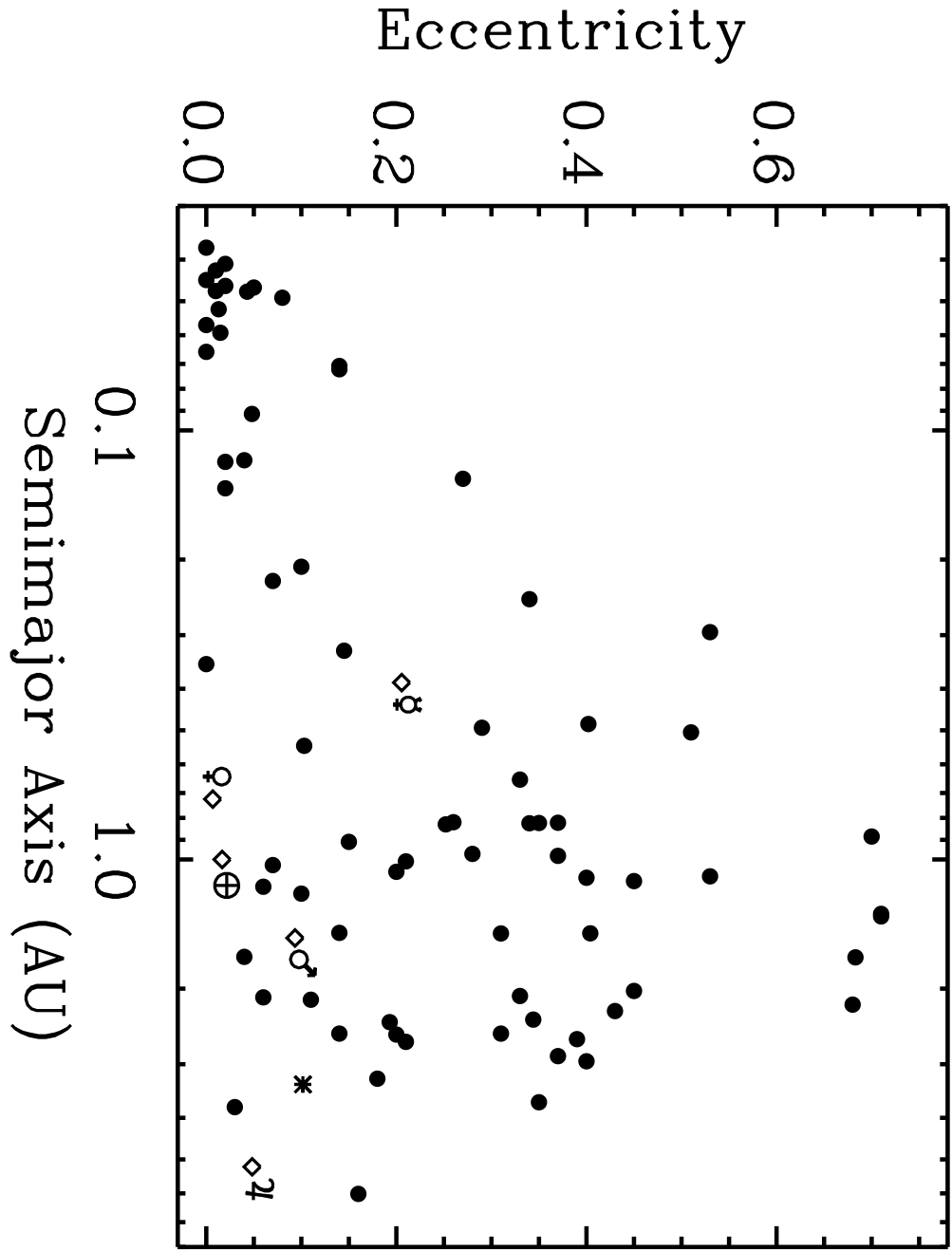}
\caption{
Orbital eccentricity vs. semimajor axis for the planet orbiting HD70642 
(asterisk), extrasolar planets listed by the IAU Working Group on 
Extrasolar Planets (filled circles), and Solar System planets out to 
Jupiter (diamonds with accompanying planetary symbols). 
HD 70642b joins 47 UMa c as the only planets
yet found in an approximately circular (e $\le$ 0.1) orbit beyond 3 AU.
}
\label{HD70642_Orbit}
\end{figure}

\end{document}